\documentclass[fleqn,twoside]{article}
\usepackage{espcrc2}

\usepackage{graphicx}
\usepackage[figuresright]{rotating}

\newcommand{\AmS}{{\protect\the\textfont2
  A\kern-.1667em\lower.5ex\hbox{M}\kern-.125emS}}
\newcommand{\eq}[1]{eq.~(\ref{#1})}

\title{
{
\vspace{-4.3cm} \normalsize \hfill
\parbox{29mm}{HU-EP-04/47\\
SFB/CPP-04-35 \\
September 2004}
}\\[30mm]
\vspace{-0.5cm}
Wilson--like fermions and the static $B_{\rm B}$ parameter with no chirality
breaking mixings
\thanks{Talk~presented at Lattice 2004, Fermi National Accelerator
  Laboratory, June 21~-~26, 2004.~Work supported by the DFG in the SFB/TR 09.}}
\author{Michele Della Morte\address[HU]{Institut f\"ur Physik, 
Humboldt--Universit\"at, Newtonstr. 15, 12489 Berlin, Germany}}
\begin{document}

\begin{abstract}
I consider the  recent proposal by R.~Frezzotti and G.~Rossi to
chirally improve Wilson fermions in such a way that mixings among operators
of different chirality can be excluded.
The method, which is based on the use of twisted mass QCD with several replica
of valence quarks, is extended here to static-light systems.
The operators relevant for the computation of the $B_{\rm B}$  parameter
(in the static approximation) are discussed. In this case the same
renormalization pattern as for Ginsparg-Wilson fermions
is obtained by a simple modification of the discretization of the
action for valence quarks.
\vspace{0.031cm}
\end{abstract}

\maketitle

\section{INTRODUCTION}
The $B_{\rm B}$ parameter describes $B -  \overline{B}$
oscillations and it is an important quantity in the analyses of
the CKM unitarity triangle. It is defined through the matrix element 
(between the states $B$ and $\overline{B}$) 
of the $\Delta b=2$ effective weak
 Hamiltonian operator $O_{\rm VV + AA}$. Here I adopt the notation 
\begin{equation}
O_{\rm YZ} = (\overline{b} \Gamma_{\rm Y} q) (\overline{b} 
\Gamma_{\rm  Z}q) \;,
\end{equation}
where $q$ denotes the light (down or strange) quark and $\Gamma_{\rm X} =
{\bf 1}, \gamma_5, \gamma_{\mu},\gamma_{\mu} \gamma_5$ for X=
  S, P, ${\rm V}_{\mu}$, ${\rm A}_{\mu}$ respectively.
  A precise theoretical prediction for $B_{\rm B}$ would provide a
  stringent test of the Standard Model. Given the definition in terms
  of an hadronic matrix element, this prediction has to be
  non--perturbative, e.g. from the lattice.

Heavy-light systems intrinsically involve scales differing by several
orders of magnitude. This problem hampers direct simulations on the
lattice. An alternative approach consists in the use of an effective
theory like HQET. It is derived through a formal expansion of the QCD
Lagrangian in powers of $1/m_b$. At the lowest order the heavy ($b$)
quark is treated as static. Analogously, operators in the full theory
(QCD) are expanded in powers of $1/m_b$, in particular
\begin{eqnarray}
&&O_{\rm VV+AA}^{\rm QCD}(m_b)= C_{\rm L} (m_b,\mu)\;  O_{\rm VV+AA}^{\rm
  HQET}(\mu) \nonumber \\ 
&&\;\;\;\;+ \;C_{\rm S}(m_b,\mu) \; O_{\rm SS+PP}^{\rm HQET} (\mu) + 
{\rm  O} (1/m_b) \; ,
\end{eqnarray}
where $\mu$ is the renormalization scale in the HQET. The functions
$C_{\rm L}(m_b, \mu)$ and $C_{\rm S}(m_b, \mu)$ have been computed at
NLO in the $\overline{\rm MS}$ scheme in~\cite{Bbmatch}. 
\section{LATTICE ACTIONS}
The lattice discretizations of the static quark action are all derived 
from the Eichten-Hill action~\cite{EH}
\begin{eqnarray}
\;\;\;\;\;\;S_{\rm stat}=\sum_x \left[ \bar{h}^{(+)}(x)\nabla_0^* \right. && 
\!\!\!\!\!\!\!\!\!\!\!
h^{(+)}(x) 
\nonumber \\
&&\!\!\!\!\!\!\!\!\!\!\!\!\!\!\!\!\!\!\!\!\!\!\!\!
\left. +\bar{h}^{(-)}(x) \nabla_0 h^{(-)}(x)\right]
\label{EHeq}
\end{eqnarray}
where $\nabla_0^*$ and $\nabla_0$ are the covariant backward and
forward derivatives respectively. The field $h^{(+)}$ annihilates a
static quark, whereas $h^{(-)}$ creates a static anti-quark. They
satisfy the constraints
\begin{equation}
{{1+\gamma_0}\over{2}} h^{(+)}=h^{(+)} \; ,\quad \quad
{{1-\gamma_0}\over{2}} h^{(-)}=h^{(-)} \;. 
\end{equation}
For the action in~\eq{EHeq} the heavy quark spin symmetry (HQS) and the
local conservation of heavy quark flavor number are realized at finite
lattice spacing. HQS in particular played an important r\^ole in
discussing the mixing pattern of the operator $O_{\rm VV + AA}$ on the
lattice \cite{Bec_Rey}. It is the invariance of the action under the
SU(2) rotations
\begin{equation}
h^{(\pm)} \rightarrow V(\vec{\phi}^{\;(\pm)}) h^{(\pm)} \; , \quad \!\!\!
\bar{h}^{(\pm)} \rightarrow \bar{h}^{(\pm)}  V(\vec{\phi}^{\;(\pm)}) \; ,
\end{equation}
with $V=\exp(-i \phi_i \varepsilon_{ijk} \sigma_{jk})$, and
transformation parameters $\phi_i$. Concerning rotational invariance,
only discrete spatial rotations remain symmetries of the static action.
The same set of symmetries is preserved by the statistically improved 
static actions proposed in~\cite{stat_imp}. 
The following discussion goes through unchanged if
those actions are used.

Moving to  the action for the light quarks, the
renormalization of the operator $O_{\rm  VV+AA}$ has been discussed 
in~\cite{Bec_Rey} for the Wilson action and for Overlap fermions~\cite{Neu}.
The latter fulfil  the Ginsparg-Wilson relation and they therefore exhibit
an exact chiral symmetry on the lattice. To fix the notation,
correlation functions of the type
\begin{eqnarray}
&& \!\!\!\!\!\!\!\!\!\!\
C_{\cal{BOB}}(x,y)= 
\nonumber \\
&& \!\!\!
\langle(\bar{q}\gamma_5 h^{(+)})(x)
O_{{\rm YZ} \pm {\rm KJ}}(0)
(\bar{q}\gamma_5h^{(-)})(y)\rangle\; ,
\end{eqnarray}
with 
$O_{{\rm YZ}}=(\bar{h}^{(+)} \Gamma_{\rm Y} q) (\bar{h}^{(-)}
  \Gamma_{\rm Z} q)$  ,
will be considered, as they provide the relevant matrix elements to
compute $B_{\rm B}$ in the static approximation.

Considering a basis of parity even $\Delta b=2$ operators:
$\{ O_{\rm VV+AA}, O_{\rm SS+PP}, O_{\rm VV-AA}, O_{\rm SS-PP}\}$, the
main result in~\cite{Bec_Rey} is that for Wilson fermions HQS and O(3)
symmetries constrain the mixings under renormalization  in this basis 
to be described by the matrix $Z$
\begin{equation}
Z\!=\!
\left(\matrix{ {Z_{11}} & 0 & {{Z}_{13}} & 2\,{{Z}_{13}} \cr 
        \frac{{Z_{11}} -{Z_{22}}}{4} & {Z_{22}} & {{Z }_{23}} &
	-{{\!Z}_{13}}\!- 2\,{{Z }_{23}} \cr 
        {{Z}_{31}} & {{Z }_{32}} & {Z_{33}} & {Z_{34}} \cr
        \frac{2\,{{Z }_{31}} - {{Z}_{32}}}{4} &
        \frac{-{{Z }_{32}}}{2} & \frac{{Z_{34}}}{4} &
        {Z_{33}} \cr  }  \right)\!\!,  
\label{mixW}
\end{equation}
whereas for Overlap fermions chiral symmetry rules out the mixings
among operators of different chirality, yielding
\begin{equation}
Z=\left(\matrix{ {Z_{11}} & 0 & 0 & 0 \cr \frac{{Z_{11}} - {Z_{22}}}
   {4} & {Z_{22}} & 0 & 0 \cr 0 & 0& {Z_{33}} & {Z_{34}} \cr
    0 & 0 & \frac{{Z_{34}}}{4} &
    {Z_{33}} \cr  }\right)\;.
\label{ZGW}
\end{equation}
\section{tmQCD} 
\vspace{-0.123cm}
Twisted mass QCD (tmQCD) has been introduced in~\cite{tmQCDI}, where it has
been proven to be a legal regularization for QCD with two
degenerate flavors. Choosing the twisting angle $\omega=\pi/2$, in the
{\it physical basis} the fermionic action for a doublet of quarks
$\psi=\left(\!\! \begin{array}{c} q\\ q' \end{array} \!\!\right)$ reads
\begin{eqnarray}
&&\!\!\!\!\!S_{\rm tm}= \sum_{\rm x} \overline{\psi}(x) \left[
{{1}\over{2}} \gamma_{\mu} (\nabla^*_{\mu} + \nabla_{\mu})+ \right. \nonumber
\\
&&\left. i\gamma_5\tau_3
\left( {{r}\over{2}} \nabla^*_{\mu}\nabla_{\mu}-M_{\rm cr}(r)
\right)+m_q \right]\psi(x)  \; ,
\label{Stm}
\end{eqnarray}
and it looks like a simple modification of the Wilson action, it just amounts
to chirally twisting the Wilson term. In~\eq{Stm} $M_{\rm cr}$ is the
usual counter-term due to the Wilson term while $m_q$ is the bare,
multiplicatively renormalizable, quark mass. It is easy to see that
in the case $m_q=0$ the action in~\eq{Stm} is invariant under 
axial transformations with generators $\tau_1$ and $\tau_2$, while 
 axial rotations generated by $\tau_3$ change the action by cutoff
effects. In particular the massless action is invariant under the finite chiral
rotations
\begin{eqnarray}
\,\,\,\,\psi \rightarrow i \gamma_5 \tau_1 \psi \quad &&{\rm or} \;\;
\tau_1 \rightarrow \tau_2  \nonumber \;,\\
\,\,\,\,\overline{\psi} \rightarrow i \overline{\psi} \gamma_5 \tau_1
\quad &&{\rm or} \;\;  \tau_1 \rightarrow \tau_2 \; .
\end{eqnarray}
On the other hand vector-flavor symmetry is in general broken, 
only $\tau_3$--vector
rotations are exactly conserved. Thus this simple modification of
Wilson regularization doesn't change the number of exactly conserved
vector/axial transformations, which is 3 in both cases. The
consequences of flavor symmetry breaking have been theoretically
investigated in $\chi$PT in~\cite{Muen,Gigi,Sharpe} while a numerical
study for example of the splitting between $\pi^0$ and $\pi^{\pm}$ is
still missing.

Concerning renormalization, having a subset of the
chiral symmetry exactly preserved should simplify the mixings. To show
that for $B_{\rm B}$ in the static approximation this is really the case 
I closely follow~\cite{impII}.  There it has been shown that using
tmQCD with several replica of the valence quarks the chirality
breaking mixings for a large set of 4-fermion operators can be ruled out. 
The exact number of replica to be introduced depends in general on the quantity
of interest. We will see that the case I'm discussing here turns out
to be among the simplest ones, as only the action for one doublet as
in~\eq{Stm} needs to be considered.
As rotational O(3) invariance and HQS are preserved by the twisting, 
the starting point is the matrix $Z$ in~\eq{mixW}. For the moment I
focus on the matrix element of  the operator $O_{\rm VV+AA}$. 
Making use of  Wick's theorem one can show that the same 
(up to cutoff effects) 
matrix element can be extracted from the correlation function
\begin{eqnarray}
&& \!\!\!\!\!\!\!\!\!\!\
C_{\cal{BQB}}(x,y)= 
\nonumber \\
&& \!\!\!
\langle(\bar{q}'\gamma_5 h^{(+)})(x)
Q_{{\rm YZ} \pm {\rm KJ}}(0)
(\bar{q}\gamma_5h^{(-)})(y)\rangle\; ,
\end{eqnarray}
with $Q_{{\rm YZ}}$ symmetrised under $q \leftrightarrow q'$
\begin{eqnarray}
\,\,\,\,\,Q_{{\rm YZ}}\!\!\!&=&\!\!\! (\bar{h}^{(+)} \Gamma_{\rm Y} \;q)
 (\bar{h}^{(-)}  \Gamma_{\rm Z} \;q') + \nonumber \\
&&\quad\quad
(\bar{h}^{(+)} \Gamma_{\rm Y} \;q') (\bar{h}^{(-)}  \Gamma_{\rm Z} \;q) \; .
\end{eqnarray}
In addition to the mixings with $Q_{\rm VV-AA}$ and $Q_{\rm SS-PP}$, 
the operators of opposite parity $Q_{\rm VA \pm AV}$ and $Q_{\rm SP \pm PS}$
need to be considered due to the parity breaking induced by the tmQCD
action. Let's introduce the transformations:
\vspace{-0.15cm}
\begin{itemize} 
\item $Ex_5$, which already appeared in~\cite{impII}
\vspace{-0.2cm}
\begin{eqnarray}
q &\rightarrow  -i \gamma_5 q' \;, \quad\quad
  \bar{q} &\rightarrow  -i\bar{q}'\gamma_5 \;,
 \nonumber\\
q' &\rightarrow   +i \gamma_5 q \;,  \quad\quad
  \bar{q}' &\rightarrow +i\bar{q}\gamma_5 \;.
\end{eqnarray}
It maps $S_{\rm tm}(m_q)$ onto $S_{\rm tm}(-m_q)$.
\item${\cal{P}}_{\pi/2}$ ($x_P=(-{\vec{x}},x_0)$)
\vspace{-0.15cm}
\begin{eqnarray}
U_0(x) &\rightarrow& U_0(x_P)\;, \nonumber \\
U_{ k}(x) &\rightarrow& U^{\dagger}_{ k}(x_P-\hat{k}) \;,\nonumber \\ 
q(x) &\rightarrow&  i\gamma_5\gamma_0 q(x_P) \;, \nonumber \\
\bar{q}(x) &\rightarrow& i \bar{q}(x_P) \gamma_0\gamma_5 \;, \nonumber \\
h^{(\pm)}(x) &\rightarrow& \gamma_5 h^{(\pm)}(x_P)\; ,  \nonumber \\
\bar{h}^{(\pm)}(x) &\rightarrow& \bar{h}^{(\pm)}(x_P) \gamma_5\; ,
\end{eqnarray}
and analogously for $q'$ and $\bar{q}'$.
Again its effect on $S_{\rm tm}$ is  a change in the sign of $m_q$.
\item ${\cal{P}}'_{\pi/2}$, same as ${\cal{P}}_{\pi/2}$ except for
\vspace{-0.2cm}
\begin{eqnarray}
h^{(\pm)}(x) &\rightarrow& \gamma_0 h^{(\pm)}(x_P)\; ,
\nonumber \\
\bar{h}^{(\pm)}(x) &\rightarrow& \bar{h}^{(\pm)}(x_P) \gamma_0\; .
\end{eqnarray}
\end{itemize}
\vspace{-0.35cm}
The $Q$-operators have been constructed to have a definite parity under
these transformations, indeed 
$Ex_5^2={\cal{P}}_{\pi/2}^2={{\cal{P}}'}_{\pi/2}^2={\bf 1}$.
Parities are summarised in table~1. The action in~\eq{Stm} 
on the  other hand is invariant under $Ex_5 \times {\cal{P}}_{\pi/2}$ and
$Ex_5 \times {\cal{P}}_{\pi/2}'$.
Thus $Ex_5 \times {\cal{P}}_{\pi/2}$ can be used to exclude mixings
of $Q_{\rm VV+AA}$ with $Q_{\rm VV-AA}$,
$Q_{\rm SS-PP}$, $Q_{\rm AV+VA}$ and $Q_{\rm SP+PS}$,
while $Ex_5 \times {\cal{P}}_{\pi/2}'$ rules out $Q_{\rm AV-VA}$
and $Q_{\rm SP-PS}$. The arguments can be repeated for the operator
$Q_{\rm SS+PP}$, again the result is that its renormalization
pattern is the same as for Overlap fermions, i.e. described by the matrix 
in~\eq{ZGW}.
\begin{table}[htb]
\begin{tabular}{cccccc}
\hline
 &  $Ex_5$ & ${\cal{P}}_{\pi/2}$   & ${\cal{P}}_{\pi/2}'$  & 
 $Ex_5 \times$ & $Ex_5 \times$\\
 &         &                       &                       &
 ${\cal{P}}_{\pi/2}$ & ${\cal{P}}_{\pi/2}'$ \\
\hline
 $Q_{\rm VV+AA}$ & even    &  even                   & odd  & {\bf even} & {\bf odd} \\
 $Q_{\rm VV-AA}$ & odd     &  even                   & even & {\bf odd} & odd \\
 $Q_{\rm SS-PP}$ & odd     &  even                   & even &{\bf odd} & odd\\
 $Q_{\rm AV+VA}$ & even    &  odd                    & even &{\bf odd}  & even\\
 $Q_{\rm AV-VA}$ & odd     &  odd                    & odd  & even & {\bf even} \\
 $Q_{\rm SP+PS}$ & even    &  odd                    & even & {\bf odd} & even\\
 $Q_{\rm SP-PS}$ & odd     &  odd                    & odd  & even & {\bf even}\\
\hline
\end{tabular}

\vspace{0.15cm}
Table~1. {\sl Parities of $Q$-operators.}
\label{pary}
\vspace{-0.7cm}
\end{table}

Finally, the renormalization factors of the operators can be computed
non-perturbatively in the Schr\"odinger functional (SF) scheme. To this
purpose it is convenient to perform the change of variables
\begin{equation}
\psi= e ^{i\frac{\pi}{2} \gamma_5 \frac{\tau_3}{2}} \chi \; , \quad
\overline{\psi}= \overline{\chi} e ^{i\frac{\pi}{2} \gamma_5 \frac{\tau_3}{2}} \; ,
\end{equation}
the twisting then completely moves to the mass term and
the action in terms of $\overline{\chi}$ and $\chi$ is consistent with
SF boundary conditions. At the same time  the operators need to be
rotated, in particular  $Q_{\rm
  VV+AA}(\bar{h}^{(\pm)},\psi) \rightarrow  Q_{\rm VV+AA}(\bar{h}^{(\pm)},\chi)$.
The approach is very
similar to the one used in~\cite{Bk,ren4}  to compute $B_{\rm K}$ and its renormalization
factor. 

\vspace{0.15cm}
{\bf Acknowledgements.} I wish to thank A.~Shindler, M.~Papinutto and
R.~Frezzotti for useful discussions.

\end{document}